\documentclass[journal]{IEEEtran}
\usepackage{caption}
\usepackage{graphicx}
\usepackage{comment}
\usepackage{tikz}

\usepackage{pgfplots,pgfplotstable}
\usetikzlibrary{pgfplots.groupplots}
\pgfplotsset{compat=1.5}

\usepackage{subcaption} 

\usepackage{multirow}
\usepackage{booktabs}
\usepackage{makecell}
\usepackage{amsmath}

\usepackage{textcomp}
\usetikzlibrary{shapes,arrows}

\usepackage{empheq} 

\usepackage{stfloats} 

\usepackage[hyphens]{url}
\usepackage[hidelinks]{hyperref}
\hypersetup{colorlinks=true,linkcolor=black,citecolor=black,filecolor=black,urlcolor=black,breaklinks=true}
\usepackage{cite} 

\usepackage{caption} 
\usepackage{color}

\usepackage{cancel} 

\usepackage{array} 

\usepackage[export]{adjustbox} 

\usepackage[normalem]{ulem} 

\usepackage{amsfonts} 

\usepackage{cite}
\usepackage[hyphens]{url}
\usepackage[hidelinks]{hyperref}
\hypersetup{colorlinks=true,linkcolor=black,citecolor=black,filecolor=black,urlcolor=black,breaklinks=true}
\usepackage{float}
\usepackage{color}
\usepackage{amsmath,amssymb,amsfonts}
\usepackage{algorithmic}
\usepackage{graphicx}
\usepackage{svg}
\usepackage{amssymb} 
\usepackage{bbding} 
\usepackage{balance}
\usepackage{textcomp}
\usepackage{xcolor}
\usepackage{multirow}
\usepackage{multicol}
\def\BibTeX{{\rm B\kern-.05em{\sc i\kern-.025em b}\kern-.08em
    T\kern-.1667em\lower.7ex\hbox{E}\kern-.125emX}}

\makeatletter
\def\bstctlcite{\@ifnextchar[{\@bstctlcite}{\@bstctlcite[@auxout]}}
\def\@bstctlcite[#1]#2{\@bsphack
 \@for\@citeb:=#2\do{%
   \edef\@citeb{\expandafter\@firstofone\@citeb}%
   \if@filesw\immediate\write\csname #1\endcsname{\string\citation{\@citeb}}\fi}%
 \@esphack}
\makeatother

\begin{document}
\bstctlcite{BSTcontrol}

\title{Voltage Unbalance-Aware AC Optimal\\Power Flow in Distribution Networks}

\author{Alireza~Zabihi,~\IEEEmembership{Student Member,~IEEE,}
        Luis~Badesa,
        and~Araceli~Hernández,~\IEEEmembership{Senior Member,~IEEE} 
}

\maketitle

\begin{abstract}
The increasing penetration of single-phase loads and distributed generation exacerbates voltage unbalance (VU) in distribution grids, raising concerns about power quality and complicating network operation. However, most market-clearing models and price-based coordination frameworks do not enforce VU limits within a three-phase AC representation, so the implications for grid-code compliance, numerical scalability, and economic signals remain unclear. This paper embeds VU in a three-phase AC optimal power flow market-clearing model and benchmarks two treatments: strict VU limit enforcement and objective function penalization. Building on these insights, an Improved Hybrid Limits (IHL) formulation is proposed that preserves compliance while using a smooth unbalance proxy in the objective to guide the optimization solver. Case studies on a European low-voltage feeder show that IHL maintains feasible operating points, yields price and curtailment signals consistent with conventional hybrid formulations, and converges substantially faster and more reliably than a penalization based on the exact unbalance metric. These results support IHL as a practical and scalable mechanism for VU mitigation in market-based operation of unbalanced distribution systems.
\end{abstract}


\IEEEpeerreviewmaketitle

\section*{Nomenclature}
\addcontentsline{toc}{section}{Nomenclature}
\subsection*{Acronyms}
\begin{IEEEdescription}[\IEEEusemathlabelsep\IEEEsetlabelwidth{MPVUR}]
    \item[AC-OPF] Three-phase AC optimal power flow
    \item[CCoG] Curtailment cost of generation
    \item[DER] Distributed energy resource
    \item[DLMP] Distribution locational marginal price
    \item[DSO] Distribution system operator
    \item[EV] Electric vehicle
    \item[IHL] Improved Hybrid Limits
    \item[KKT] Karush--Kuhn--Tucker conditions
    \item[LV] Low voltage
    \item[MPVUR] Modified phase voltage unbalance ratio
    \item[OPF] Optimal power flow
    \item[PV] Photovoltaics
    \item[PVUR] Phase voltage unbalance ratio
    \item[VU] Voltage unbalance
    \item[VUF] Voltage unbalance factor
\end{IEEEdescription}    

\subsection*{Indices and Sets}
\begin{IEEEdescription}[\IEEEusemathlabelsep\IEEEsetlabelwidth{$\varphi,\,\Phi$}]
    \item[$i,\,\mathcal{N}$] Index, Set of three-phase nodes.
    \item[$\varphi,\,\Phi$] Index, Set of electrical phases.
    \item[$\theta_i^p$] Set of three-phase voltage phase angles of node $i$.
    \vspace*{3pt}
    \item[$v_i^{3\varphi}$] Set of three-phase voltages of node $i$.
    \item[$V_i^m$] Set of three-phase voltage magnitudes of node $i$.
\end{IEEEdescription}

\vspace*{-6pt}
\subsection*{Constants and Functions}
\begin{IEEEdescription}[\IEEEusemathlabelsep\IEEEsetlabelwidth{$g(v_i^{3\varphi})$}]
    \item[$\alpha$] Tuning coefficient of VU penalty reflecting damage cost of VU (scalar).
    \item[$C_{i_{\varphi}}^c$] Utility function of consumers connected to node $i$ on phase $\varphi$ (€/h).
    \item[$C_{i_{\varphi}}^g$] Cost function of generators connected to node $i$ on phase $\varphi$ (€/h).
    \item[$f(v_i^{3\varphi})$] Function defining the VU metric in the grid as a constraint.
    \item[$g(v_i^{3\varphi})$] Function defining the VU metric in the grid as a penalty term.
    \item[$\overline{V.U.}$] Maximum admissible level of VU.
\end{IEEEdescription}

\subsection*{Primal Variables}
\begin{IEEEdescription}[\IEEEusemathlabelsep\IEEEsetlabelwidth{$P_{i_{\varphi}}^{g}$}]
    \item[$P_{i_{\varphi}}^c$] Active power consumption at node $i$, phase $\varphi$ (kW).
    \item[$P_{i_{\varphi}}^g$] Active power generation at node $i$, phase $\varphi$ (kW).
\end{IEEEdescription}

\subsection*{Metric Variables}
\begin{IEEEdescription}[\IEEEusemathlabelsep\IEEEsetlabelwidth{$v_{\varphi}$}]
    \item[$\delta_{\varphi}$] Voltage angle deviation of phase $\varphi$ at a node (rad).
    \item[$N$] Negative-sequence voltage component of the three-phase voltage at a node (p.u.).
    \item[$P$] Positive-sequence voltage component of the three-phase voltage at a node (p.u.).
    \item[$\theta$] Nominal phase angle of three-phase system (rad).
    \item[$V_{\varphi}$] Voltage magnitude of phase $\varphi$ at a node (p.u.).
    \item[$v^{\varphi}$] Voltage phasor of phase $\varphi$ at a node (phasor).
    \item[$Z$] Zero-sequence voltage component of the three-phase voltage at a node (p.u.).
\end{IEEEdescription}

\section{Introduction}
\label{sec:introduction}
\IEEEPARstart{T}{he} rapid proliferation of Distributed Energy Resources (DER)—including rooftop photovoltaics (PV) and Electric Vehicles (EVs)—is shifting power systems from centralized paradigms toward distribution-level coordination with highly heterogeneous, often single-phase, flexible assets~\cite{MPC}. In response, local electricity markets have been proposed to schedule DER, enable peer-to-peer trading, and procure grid services while maximizing social welfare~\cite{Hierarchical}. Regulatory initiatives such as FERC Order 2222 further reinforce this trajectory by enabling aggregated DER participation in wholesale markets~\cite{FERC2222}. These developments elevate a central requirement for distribution-level market design: economic efficiency and renewable integration must be pursued without compromising technical operating limits, which are tighter and more phase-dependent in low-voltage networks than in transmission systems.

Among the most consequential power quality constraints in modern distribution grids is Voltage Unbalance (VU), which arises when phase voltages deviate from equality in magnitude and/or the ideal 120$^\circ$ symmetry in angle. VU produces negative-sequence components that increase losses and thermal stress in rotating machines and transformers, accelerating insulation aging and reducing equipment lifetime~\cite{VU-effect}. Importantly, the impact is not merely qualitative: empirical evidence indicates that even small VU changes can materially affect induction motor lifetime, with reported sensitivities on the order of one year reduction per 0.2\% VU increment~\cite{Jalilian-Induction, Molzahn_rep}. In low-voltage feeders, VU is largely driven by uneven phase loading and is increasingly exacerbated by single-phase DER (e.g., residential PV and EV chargers) that introduce phase-asymmetric injections and consumption~\cite{CVU-DS}. As DER penetration grows, the Voltage Unbalance Factor (VUF) can exceed recommended thresholds~\cite{Analysis-Rich_PV_EV}, raising both compliance and reliability concerns. VUF is the most widely accepted index for measuring voltage unbalance and is defined as the ratio of the negative- to positive-sequence components~\cite{IEC61000_3_13}. 

At the same time, controllable DER interfaces (notably PV inverters) can provide voltage support and phase-balancing services when suitably coordinated, creating an opportunity to mitigate VU through market-compatible control signals~\cite{Rahul_Gupta}. The operational flexibility of distribution networks is constrained by regulatory frameworks, as grid codes typically impose strict voltage unbalance limits to protect network assets and end-user equipment. A comparison of these limits across various international standards and regional grid codes is summarized in Table~\ref{tab:voltage_standards}. These requirements typically cap the VU level between 1\% and 3\%, depending on the jurisdiction and the specific voltage level of the point of common coupling~\cite{IEC61000_3_13, EN50160, ANSI_C84_1, IEEE1159, NEMA_MG1, UK_P29}. Consequently, any distribution-level market-clearing mechanism that materially changes phase-wise injections must either (i) explicitly enforce VU limits or (ii) provide provable safeguards that the resulting operating point remains compliant.

\begin{table}[!t]
\renewcommand{\arraystretch}{1.3}
\centering
\caption{Comparison of Voltage Unbalance Limits Across Different Standards and Grid Codes.}
\begin{tabular}{lcc}
\toprule
\textbf{Standard / Code} & 
\makecell{\textbf{Voltage}\\\textbf{Level}} & 
\makecell{\textbf{Recommended}\\\textbf{Limit (\%)}} \\
\midrule
IEC 61000-3-13 & MV/HV/EHV & 2.0 \\
EN 50160       & LV/MV      & 2.0 -- 3.0 \\
ANSI C84.1     & LV/MV      & 3.0 \\
IEEE 1159      & All        & 3.0 \\
NEMA MG1       & LV         & 1.0 -- 5.0 \\
UK P29         & LV/MV      & 1.3 -- 2.0 \\
\bottomrule
\end{tabular}
\label{tab:voltage_standards}
\end{table}

Recent work has therefore advanced three-phase AC Optimal Power Flow (AC-OPF) formulations that internalize unbalance-related physics and controls. Representative approaches include minimizing VUF via reactive power control of inverter-interfaced DER~\cite{Line-std} and enforcing inequality constraints on negative-sequence voltage magnitude or directly on VUF within the OPF~\cite{Op-Inv-Volt-profile}. However, the exact VUF constraint is nonlinear and non-convex because it involves ratios of sequence components and, in general, requires sensitivities of negative-sequence quantities with respect to control variables~\cite{churkin2024}. This modeling burden often leads to tractability-driven approximations (e.g., linearization, bounding, or surrogate constraints) to obtain numerically reliable solutions at scale~\cite{wang2021three}. As a result, the current state of the art reflects a persistent tension between physical fidelity (accurate, enforceable unbalance limits) and computational practicality (robust convergence and scalability in three-phase non-convex AC-OPF).

A key design choice underlying this tension is whether VU requirements are treated as \textit{hard limits} (explicit constraints) or \textit{soft limits} (penalty terms in the objective function). Hard limits offer deterministic compliance and equipment protection, but they can amplify non-convexity and slow convergence in unbalanced AC-OPF, motivating additional relaxations in the literature~\cite{ACOPF-MPhase, Investig-nonconv}. Soft limits, by contrast, enlarge the feasible region and can improve numerical behavior by allowing controlled violations; they can also represent an economic proxy for the degradation and loss impacts of VU. Yet, soft limits alone risk producing solutions that violate grid-code thresholds if penalty weights are not tuned appropriately~\cite{Line-std, MY-J1}. This \textit{hard}--\textit{soft} trade-off is consistent with classical optimization viewpoints: penalty methods enforce requirements through increasing costs for violations, whereas barrier-like mechanisms impose steep costs near boundaries to discourage infeasibility~\cite{Penalty&Barrier}. In practice, penalty tuning and robustness remain nontrivial, particularly in distribution OPF settings where binding constraints shift across phases and operating points~\cite{Opt-Constraint, Modified-Barrier}.

From a numerical standpoint, hard VUF constraints can markedly deteriorate convergence in three-phase AC-OPF by adding sequence-component mappings and ratio-type nonlinearities that intensify phase coupling. Tightly binding unbalance constraints may induce degeneracy or near linear dependence among active constraints, producing an ill-conditioned KKT system and unreliable Newton directions~\cite{NumOpt2006}. Stringent bounds can also shrink the feasible region and weaken interior-point performance due to the reliance on strictly interior iterates, while poor scaling and frequent activation of restrictive limits further degrade conditioning~\cite{Ipopt2006, NumOpt2006}.

Despite progress in OPF-based VU mitigation, embedding unbalance-aware formulations into market clearing remains comparatively underdeveloped. Most local market designs prioritize energy balance and economic objectives and rely on simplified network representations, operating envelopes, or decomposition strategies to ensure security~\cite{Saeed, Clear-Mech, wang2022dlmp}. For example, dynamic operating envelopes provide admissible DER ranges consistent with grid limits without solving a full OPF at every dispatch interval~\cite{Nando_Ochoa, churkin2024}. Likewise, peer-to-peer and hierarchical market algorithms often focus on trading efficiency and price discovery but rarely incorporate explicit multi-phase unbalance constraints within the clearing optimization~\cite{Hierarchical}. Consequently, a clear research gap persists: market mechanisms that jointly optimize economic outcomes and phase unbalance using a three-phase AC model, while retaining numerical tractability and regulatory compliance, remain limited.

This gap is particularly salient because distribution-level economic signals are inherently phase- and location-dependent. Distribution Locational Marginal Pricing (DLMP) generalizes the LMP concept to unbalanced distribution networks, producing nodal--phase price components that reflect losses, congestion, and voltage-related scarcity~\cite{wang2022dlmp}. In parallel, the Curtailment Cost of Generators (CCoG) quantifies the opportunity cost associated with restricting DER output~\cite{APC}. Together, DLMP and CCoG provide a principled lens to evaluate how different VU-integration strategies shift both incentives and welfare, yet existing market-clearing formulations often omit explicit VU constraints, leaving a disconnect between economic optimality and power quality compliance.

To address this disconnect, this paper proposes an Improved Hybrid Limits (IHL) formulation for VU-aware distribution-level market clearing. The hybrid limits concept combines the compliance assurance of hard VUF bounds with the numerical flexibility of a penalty in the objective function. IHL strengthens this concept by introducing a smooth surrogate penalty index derived from the relationship between the Phase Voltage Unbalance Ratio (PVUR) --a standard VU index defined by IEEE in~\cite{PVUR2-112} and~\cite{PVUR2-936}-- and VUF, while preserving hard VUF bounds to maintain grid-code compliance. Embedding IHL within a three-phase AC-OPF enables scalable market clearing that yields economically meaningful DLMPs and CCoGs under explicit VU considerations.

The main contributions of this work are:
\begin{enumerate}
\item A new VU-aware market-clearing formulation based on a three-phase AC-OPF with Improved Hybrid Limits (IHL), combining hard VUF compliance with a smooth unbalance penalty to improve numerical tractability and scalability.
\item Derivation of an optimization‑amenable surrogate unbalance index (via a modified PVUR--VUF relationship) suitable for inclusion as a differentiable penalty term within the AC-OPF objective function.
\item A systematic comparative assessment framework using DLMP and CCoG to quantify how hard, soft, and hybrid VU treatments reshape economic signals and curtailment impacts in distribution-level electricity markets.
\end{enumerate}

The remainder of this paper is organized as follows: Section~\ref{sec:methodology} presents the mathematical formulation, including the surrogate function and its properties; Section~\ref{sec:test} reports the case studies and comparative results; and Section~\ref{sec:conclusion} concludes the paper.

\section{Methodology}
\label{sec:methodology}
This section introduces the hybrid limits concept for incorporating voltage unbalance into three-phase AC-OPF market clearing. It then derives a modified index and a correction coefficient using the relationship between VUF and PVUR. Finally, it presents a sensitivity analysis to show how the modified index behaves relative to VUF and to highlight the resulting numerical simplicity and practical benefits.

\subsection{Hybrid Methodologies}
The proposed hybrid methodology incorporates a bus-level VU index as a penalty term in the objective function (a typical `soft limit' strategy), while simultaneously enforcing VUF through an explicit constraint (hard limit). This dual-layer structure preserves compliance with grid-code limits and, at the same time, increases numerical flexibility by allowing the optimizer to trade off economic objectives against a penalized (and controllable) unbalance measure.

For clarity, a simplified social welfare maximization problem is stated as follows:
\begin{equation}
\label{eq:social_welfare2}
\begin{gathered}
\max \sum_{\substack{i \in \mathcal{N} \\
\varphi \in \Phi}} C_{i_{\varphi}}^c\left(P_{i_{\varphi}}^c\right)-C_{i_{\varphi}}^g\left(P_{i_{\varphi}}^g\right)-\alpha \cdot g\left(v_i^{3\varphi}\right) \\
\end{gathered}
\end{equation}
\vspace{-4mm}
\begin{equation}
\label{eq:network}
\begin{gathered}
\text{Network constraints}
\end{gathered}
\end{equation}
\begin{equation}
\label{eq:V.U._constraint}
\begin{gathered}
f\left(v_i^{3\varphi}\right) \leq \overline{V . U .}
\end{gathered}
\end{equation}

Constraints~(\ref{eq:network}) represent the operating requirements of the grid, including power balance, voltage bounds, generation limits, and other system constraints (a detailed implementation of the full model is discussed in the Section~\ref{sec:test}). Constraint~(\ref{eq:V.U._constraint}) enforces the hard VU limit required by the applicable grid codes.

Several special cases follow directly. By appropriate choices of 
\( f \) and \( g \), the proposed framework recovers several special cases of practical interest. If \( g = f = 0 \), the formulation reduces to a basic OPF without VU considerations. If only \( f \) is imposed, the formulation corresponds to \emph{hard limits}. If only \( g \) is included, the formulation corresponds to \emph{soft limits}. When both \( g \) and \( f \) are imposed and \( g = f = \text{VUF} \), the formulation represents the basic `hybrid limits' method. Finally, when \( f \) remains as VUF but a simpler surrogate function is adopted for \( g \), the resulting formulation is referred to as the Improved Hybrid Limits method.

\subsection{Modified PVUR}
\label{MPVUR}
Several standards define acceptable VU metrics and levels~\cite{MY-PT2025}. The simplest definition, in terms of required data and computational effort, is specified by IEEE in~\cite{PVUR2-112} and~\cite{PVUR2-936} and is commonly referred to as PVUR. This index is computed using the average phase-voltage magnitude and the spread between the maximum and minimum phase-voltage magnitudes, as follows:
\begin{equation}
V_{\text{avg}}^P = \frac{V_a + V_b + V_c}{3}
\label{eq:phase_avg_voltage_2}
\end{equation}
\begin{equation}
V_{\text{max}} = \max\left(V_a,V_b,V_c\right)
\label{eq:phase_max_voltage}
\end{equation}
\begin{equation}
V_{\text{min}} = \min\left(V_a,V_b,V_c\right)
\label{eq:phase_min_voltage}
\end{equation}

PVUR is calculated as:
\begin{equation}
\text{PVUR} = \frac{V_{\text{max}} - V_{\text{min}}}{V_{\text{avg}}^P} \times 100\%
\label{eq:phase_unbalance_ratio_2}
\end{equation}

To derive a corrected version of PVUR based on its relationship with VUF, the positive-, negative-, and zero-sequence components are first analyzed under normal operating conditions of the distribution grid.

The general formulation of a three-phase system in phasor representation is expressed as follows:
\begin{align}
\label{3phase_general}
    v_a &= V_a \angle (\theta + \delta_a) \nonumber \\
    v_b &= V_b \angle \left( \theta - \frac{2\pi}{3} + \delta_b \right) \nonumber \\
    v_c &= V_c \angle \left( \theta + \frac{2\pi}{3} + \delta_c \right)
\end{align}
Although VU can arise from both magnitude asymmetry and deviations from the ideal 120$^\circ$ phase separation, the proxy developed in this subsection targets the dominant contribution observed in practical LV feeders: unequal phase voltage magnitudes. In steady-state distribution operation, phase-angle measurements are typically unavailable in common metering infrastructure, and magnitude-based unbalance indicators are therefore widely used and shown to approximate the true VUF in realistic LV networks~\cite{Hashmi2022}. Moreover, while VUF depends on both voltage magnitudes and relative phase angles, magnitude-based definitions are analytically related to and can approximately bound the symmetrical-component-based VUF under typical operating ranges~\cite{Girigoudar2019}. Accordingly, for the remainder of this derivation, the phase-angle deviations are neglected by setting \(\delta_a=\delta_b=\delta_c=0\) in~(\ref{3phase_general}). The corresponding approximation error is assessed later (see Section~\ref{Accuracy of MPVUR}).

The Fortescue transformation matrix can now be applied to determine the sequence components. The symmetrical components \( v_0 \), \( v_1 \), and \( v_2 \) are calculated from the phase voltages \( v_a \), \( v_b \), and \( v_c \) as follows:
\begin{equation}
\label{eq:symmetrical_components}
\begin{bmatrix}
v_0 \\ v_1 \\ v_2
\end{bmatrix} = \frac{1}{3} 
\begin{bmatrix}
1 & 1 & 1 \\
1 & a & a^2 \\
1 & a^2 & a
\end{bmatrix}
\begin{bmatrix}
v_a \\ v_b \\ v_c
\end{bmatrix}
\end{equation}
where \( a = e^{j \frac{2\pi}{3}} \) is the 120$^\circ$ phase shift operator.
The sequence components are expressed as follows:
\begin{align}
    &v_0 = \frac{1}{3} \left[ \cos \theta \left( V_a - \frac{V_b}{2} - \frac{V_c}{2} \right) + \sin \theta \left( \frac{V_b\sqrt{3}}{2} - \frac{V_c\sqrt{3}}{2} \right) \right] \notag \\
    &+ \frac{1}{3} j \left[ \sin \theta \left( V_a - \frac{V_b}{2} - \frac{V_c}{2} \right) + \cos \theta \left( \frac{V_c\sqrt{3}}{2} - \frac{V_b\sqrt{3}}{2} \right) \right]
\end{align} 
\vspace*{-25pt}
\begin{align}
\label{pos-seq}
    v_1 &= \frac{1}{3} \left[ \cos \theta (V_a + V_b + V_c) + j \sin \theta (V_a + V_b + V_c) \right]
\end{align} 
\vspace*{-22pt}
\begin{align}
    &v_2 = \frac{1}{3} \left[ \cos \theta \left( V_a - \frac{V_b}{2} - \frac{V_c}{2} \right) + \sin \theta \left( \frac{V_c\sqrt{3}}{2} - \frac{V_b\sqrt{3}}{2} \right) \right] \notag \\
    &+ \frac{1}{3} j \left[ \sin \theta \left( V_a - \frac{V_b}{2} - \frac{V_c}{2} \right) + \cos \theta \left( \frac{V_b\sqrt{3}}{2} - \frac{V_c\sqrt{3}}{2} \right) \right]
\end{align}
From \eqref{pos-seq}, the phase angle of \( v_1 \) is equal to \( \theta \). The phase angles of \( v_0 \) and \( v_2 \) are analyzed below:

\begin{equation}
\begin{aligned}
\angle v_0 
&= \tan^{-1}\left(\frac{A\sin\theta\ + B\cos\theta\ }{A\cos\theta\ - B\sin\theta\ } \right) \\
&= \tan^{-1}\left(\frac{\tan\theta + \frac{B}{A}}{1 - \tan\theta \cdot \frac{B}{A}} \right) \\
&= \tan^{-1}\left(\tan\left(\theta + \tan^{-1}\left(\frac{B}{A}\right)\right)\right) \\
&= \theta + \tan^{-1}\left(\frac{B}{A}\right)
\end{aligned}
\end{equation}

\noindent where 
\( A = V_a - \frac{V_b + V_c}{2} \), \quad
\( B = \frac{\sqrt{3}}{2}(V_c - V_b) \). \\

By defining \( \tan^{-1}\!\left(\frac{B}{A}\right) = \beta \), the expression can be rewritten as:
\begin{equation}
\begin{aligned}
\angle v_0 & = \theta + \beta
\end{aligned}
\end{equation}

Using the same approach, it can be concluded that:
\begin{equation}
\begin{aligned}
\angle v_2 & = \theta - \beta
\end{aligned}
\end{equation}

The magnitude of \( v_1 \) is given by \( \frac{V_a + V_b + V_c}{3} \). The magnitudes of \( v_0 \) and \( v_2 \) are analyzed below:

\begin{equation}
\begin{aligned}
|v_0| &= |v_2| = \frac{1}{3} \sqrt{\left(V_a - \frac{V_b}{2} - \frac{V_c}{2}\right)^2 + \frac{3}{4}(V_b - V_c)^2}
\end{aligned}
\end{equation}

By defining \( \frac{1}{3} \sqrt{\left(V_a - \frac{V_b}{2} - \frac{V_c}{2}\right)^2 + \frac{3}{4}(V_b - V_c)^2} = X \), the expression can be rewritten as:
\begin{equation}
\begin{aligned}
|v_0| &= |v_2| = X
\end{aligned}
\end{equation}

The most important conclusions can be summarized as follows:
\begin{align}
\text{1)} \quad & v_1 = \frac{V_a + V_b + V_c}{3} \angle \theta \label{eq:V&P1} \\
\text{2)} \quad & v_0 = X \angle (\theta + \beta), \quad v_2 = X \angle (\theta - \beta) \label{eq:V&P2}
\end{align}

The values of \( X \) and \( \beta \) are computed explicitly.

For notational convenience, (\ref{eq:V&P1})--(\ref{eq:V&P2}) can be rewritten by introducing the magnitudes \(Z\), \(P\), and \(N\) as:
\begin{align}
    v_0 &= Z \angle (\theta + \beta) \nonumber \\
    v_1 &= P \angle \theta \nonumber \\
    v_2 &= N \angle (\theta - \beta)
\end{align}

\noindent where 
\( Z = N = X \), \quad
\( P = \dfrac{V_a + V_b + V_c}{3} \). \\

Since \( \theta \) represents the phase shift of phase `a' and the reference frame is considered to be zero, it can be assumed \( \theta = 0 \) without loss of generality. The above equations can then be rewritten as follows:
\begin{align}
    v_0 &= N \angle \beta \nonumber \\
    v_1 &= P \angle 0 \nonumber \\
    v_2 &= N \angle (-\beta)
\end{align}

The reverse transformation matrix can now be applied to derive the voltages of the three-phase system and their corresponding magnitudes:
\[
\begin{bmatrix}
v_a \\ v_b \\ v_c
\end{bmatrix}
=
\begin{bmatrix}
1 & 1 & 1 \\
1 & a^2 & a \\
1 & a & a^2
\end{bmatrix}
\begin{bmatrix}
v_0 \\ v_1 \\ v_2
\end{bmatrix}
,\quad a = 1 \angle \frac{2\pi}{3}
\]

\begin{align}
    |v_a| &= |2N \cos\beta + P| \\
    |v_b| &= |2N \cos\left(\beta + \frac{2\pi}{3}\right) + P| \\
    |v_c| &= |2N \cos\left(\beta - \frac{2\pi}{3}\right) + P|
\end{align}

If \( N \leq \frac{P}{2} \), the right-hand sides of the above expressions remain nonnegative. Equivalently, this condition corresponds to \(\text{VUF} \leq 50\%\), and violating it would require an exceptionally large unbalance level. Therefore, the absolute-value terms can be removed.
Now, it is possible to compute the difference between the maximum and minimum values of the three-phase voltages. To formalize this, the following lemma is introduced, which provides a general approach for determining the difference between the maximum and minimum of any three real values. It can be expressed as:
\begin{align}
\max\left(R_1,R_2,R_3\right) - \min\left(R_1,R_2,R_3\right) = \notag \\
\frac{| R_1 - R_2 | + | R_2 - R_3 | + | R_3 - R_1 |}{2} 
\end{align}

The function representing the difference between the maximum and minimum magnitudes of the three-phase voltages is denoted as \( H(N,\beta) \):
\begin{align}
H(N, \beta) &= V_{\text{max}} - V_{\text{min}} = |N \cos\beta - N \cos\left(\beta + \frac{2\pi}{3}\right)| \notag \\
&+ |N \cos\left(\beta + \frac{2\pi}{3}\right) - N \cos\left(\beta - \frac{2\pi}{3}\right)| \notag \\
&+ |N \cos\left(\beta - \frac{2\pi}{3}\right) - N \cos\beta|    
\end{align}

This is a linear function with respect to \( N \) and a periodic function with respect to \( \beta \), having a period of \( \dfrac{\pi}{3} \). The maximum and minimum values are given by:
\begin{align}
    \text{Max}[H(N, \beta)] &= 2\sqrt{3}N \quad &&(\text{at } \beta = 0) \\
    \text{Min}[H(N, \beta)] &= 3N \quad &&(\text{at } \beta = \frac{\pi}{6})
\end{align}

By factoring \( N \) from \( H(N, \beta) \), it can be expressed as \( H(N, \beta) = \text{Const.} \times N \). Based on prior knowledge~\cite{MY-PT2025} regarding PVUR and its tendency to overestimate VUF, the largest correction factor is selected. Thus, \( \text{Const.} = 2\sqrt{3} \) is introduced:
\begin{align}
\label{eq:H(n.B)}
V_{\text{max}} - V_{\text{min}} = H(N, \beta) = 2\sqrt{3} \times N
\end{align}

Now it is possible to establish the relationship between PVUR and VUF. VUF is defined as the ratio of the negative- to positive-sequence components~\cite{IEC61000_3_13}. Based on equations (\ref{eq:V&P1}) and (\ref{eq:H(n.B)}), VUF can be expressed as:
\begin{align}
\label{eq:fin}
\begin{aligned}[t]
        \text{VUF} &= \frac{N}{P} \times 100\% \\[8pt]
        &= \frac{V_{\text{max}} - V_{\text{min}}}{2\sqrt{3} \ \dfrac{V_a + V_b + V_c}{3}} \times 100\% \\[8pt]
        &= \frac{\text{PVUR}}{2\sqrt{3}} \times 100\% = \textbf{Modified PVUR}
    \end{aligned}
\end{align}

In other words, Modified PVUR (MPVUR) is a magnitude-based approximation of VUF obtained under the assumption of negligible phase-angle deviations from the ideal 120$^\circ$ symmetry. The accuracy of MPVUR is evaluated in Section~\ref{Accuracy of MPVUR} using a realistic European LV network, where the resulting VUF and MPVUR values are compared and their correlation and mean error are reported to quantify the approximation quality.

\subsection{Sensitivity analysis of VU functions}
To provide insight into the numerical behavior of VUF and MPVUR, and to clarify their impact on the optimization model and solver performance, a sensitivity analysis is conducted. This comparison makes it evident why MPVUR is easier to handle than the original VUF in the proposed IHL formulation.

The formulation of MPVUR and its derivatives with respect to voltage magnitudes is presented below. Without loss of generality, an ordering among the phase-voltage magnitudes is assumed: \( V_a \geq V_b \geq V_c \). Therefore, MPVUR can be expressed as:
\begin{equation}
\begin{aligned}
\label{eq:MPVUR}
g(V_a, V_b, V_a) = \text{MPVUR} = \frac{1}{2\sqrt{3}} \frac{V_a - V_c}{\dfrac{V_a + V_b + V_c}{3}}
\end{aligned}
\end{equation}

The sensitivity with respect to voltage magnitudes is computed as follows:

For the phase with the largest voltage magnitude:
\begin{equation}
\begin{aligned}
\frac{\partial g}{\partial V_a} &=  \frac{3}{2\sqrt{3}} \frac{2V_c + V_b}{(V_a + V_b + V_c)^2}
\end{aligned}
\end{equation}

For the middle phase:
\begin{equation}
\begin{aligned}
\frac{\partial g}{\partial V_b} &=  \frac{3}{2\sqrt{3}} \frac{V_c - V_a}{(V_a + V_b + V_c)^2}
\end{aligned}
\end{equation}

For the phase with the smallest voltage magnitude:
\begin{equation}
\begin{aligned}
\frac{\partial g}{\partial V_c} &=  \frac{3}{2\sqrt{3}} \frac{-2V_a - V_b}{(V_a + V_b + V_c)^2}
\end{aligned}
\end{equation}

As expected, increasing the magnitude of the largest phase voltage or decreasing the magnitude of the smallest phase voltage has the greatest impact on increasing the VU level, whereas variations in the middle phase have a significantly smaller effect. This analysis remains valid only as long as the assumed ordering of phase voltages is preserved; if the ordering changes, discontinuities may occur.

The above expressions are evaluated around a nominal operating point by setting the phase-voltage magnitudes to 1 p.u. and perturbing only the phase under analysis about unity.

For the phase with the largest voltage magnitude:
\begin{equation}
\label{eq:Ph_a}
\begin{aligned}
\frac{\partial g}{\partial V_a} = \frac{9}{2\sqrt{3}} \frac{1}{(V_a + 2)^2} \cong \frac{1}{2\sqrt{3}}
\end{aligned}
\end{equation}
For the middle phase:
\begin{equation}
\label{eq:Ph_b}
\begin{aligned}
\frac{\partial g}{\partial V_b} = 0
\end{aligned}
\end{equation}
For the phase with the smallest voltage magnitude:
\begin{equation}
\label{eq:Ph_c}
\begin{aligned}
\frac{\partial g}{\partial V_c} = \frac{-9}{2\sqrt{3}} \frac{1}{(2 + V_c)^2} \cong \frac{-1}{2\sqrt{3}}
\end{aligned}
\end{equation}

The formulation of VUF and its derivatives with respect to voltage magnitudes and phase angles is presented below.
Because these components are represented as phasors, the computation requires taking the square root of the product of each phasor and its complex conjugate. To simplify the calculations, the square root step is omitted and \( \text{VUF}^2 \) is used instead.
\begin{equation}
\begin{aligned}
\label{eq:VUF2}
&f(V_n^m, \theta_n^p) = \text{VUF}^2(V_n^m, \theta_n^p) = \frac{J}{K}, \ \text{in which}  \\
J &= \left[ V_a \cos(\theta_a) + V_b \cos(\theta_b - \frac{2\pi}{3}) + V_c \cos(\theta_c + \frac{2\pi}{3}) \right]^2 \\
&+ \left[ V_a \sin(\theta_a) + V_b \sin(\theta_b - \frac{2\pi}{3}) + V_c \sin(\theta_c + \frac{2\pi}{3}) \right]^2 \\
&= J^2_{1} + J^2_{2} \\
K &= \left[ V_a \cos(\theta_a) + V_b \cos(\theta_b + \frac{2\pi}{3}) + V_c \cos(\theta_c - \frac{2\pi}{3}) \right]^2 \\
&+ \left[ V_a \sin(\theta_a) + V_b \sin(\theta_b + \frac{2\pi}{3}) + V_c \sin(\theta_c - \frac{2\pi}{3}) \right]^2 \\
&= K^2_{1} + K^2_{2}
\end{aligned}
\end{equation}

In this formulation, the cosine term of \( J \) is denoted by \( J_{1} \), and the sine term is denoted by \( J_{2} \). Similarly, the corresponding terms for \( K \) are \( K_{1} \) and \( K_{2} \). To compute the sensitivity with respect to voltage magnitude and phase angle for a generic phase (here, phase `a'), the derivatives are obtained by varying only one parameter (magnitude or angle) while keeping all other variables at their nominal values:
\begin{equation}
\begin{aligned}[b]
\frac{\partial f(V_n^m, \theta_n^p)}{\partial V_a} &= \frac{2}{K^2} \big[ K(J_{1}\cos\theta_a + J_{2}\sin\theta_a) \\
&\quad - J(K_{1}\cos\theta_a + K_{2}\sin\theta_a) \big] \\
&= \frac{6(V_a -1)}{(V_a + 2)^3}
\end{aligned}
\end{equation}
\begin{equation}
\begin{aligned}[b]
\frac{\partial f(V_n^m, \theta_n^p)}{\partial \theta_a} &= \frac{2V_a}{K^2} \big[ K(J_{2}\cos\theta_a - J_{1}\sin\theta_a) \\
&\quad - J(K_{2}\cos\theta_a - K_{1}\sin\theta_a) \big] \\
&= \frac{18\sin\theta_a}{(5 + 4\cos\theta_a)^2}
\end{aligned}
\end{equation}

\begin{figure}[!t]
    \centering
    \includegraphics[width=3.2in]{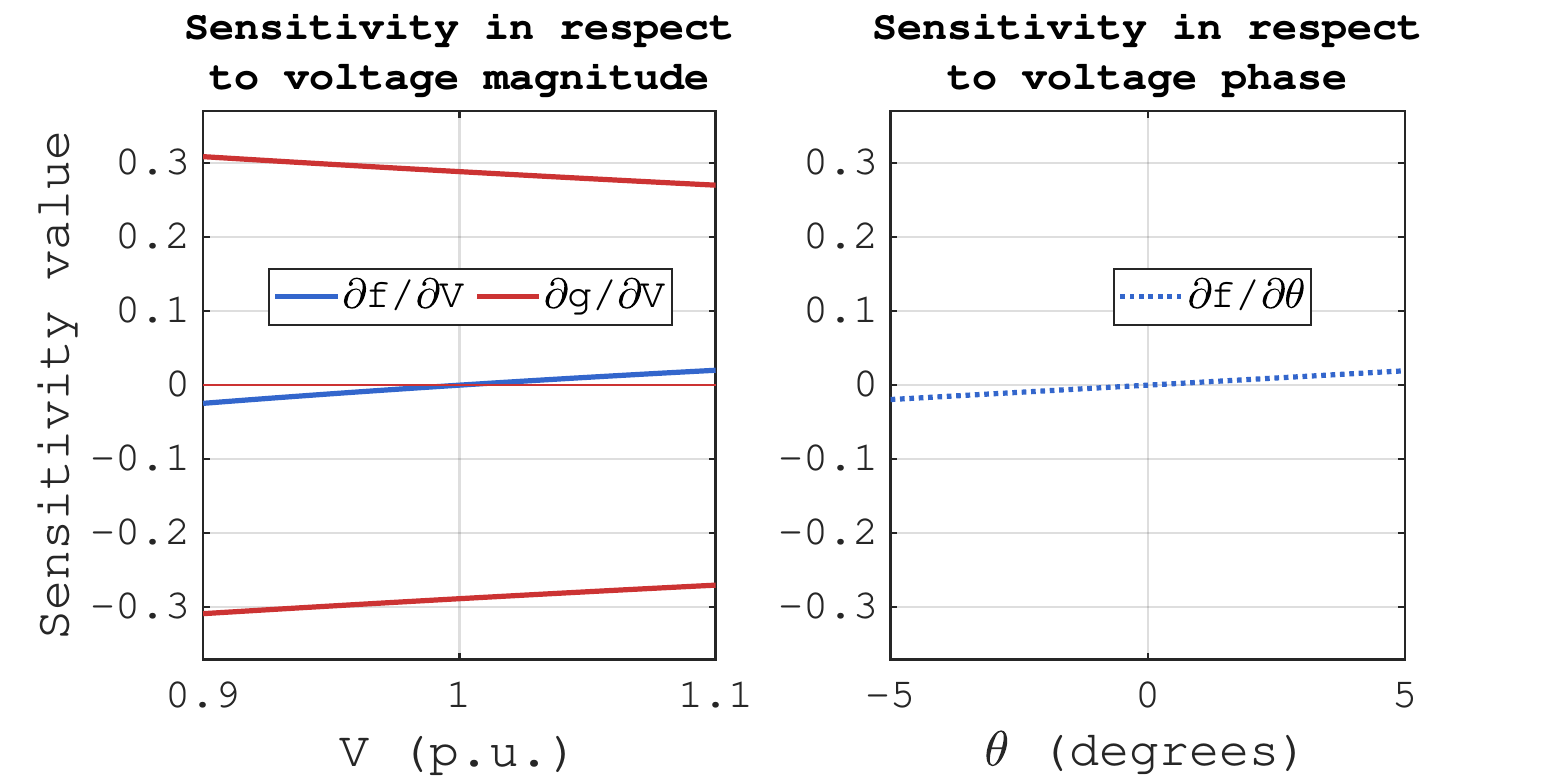}
    \caption{Sensitivity analysis of VUF(=f) and MPVUR(=g).}
    \label{fig:Sensivity analysis illustration.}
\end{figure}

By approximating $\frac{\partial f(V_n^m, \theta_n^p)}{\partial V_a}$ and $\frac{\partial f(V_n^m, \theta_n^p)}{\partial \theta_a}$ around their nominal values ($V_a=1$, $\theta_a=0$), it can be concluded that:
\begin{equation}
\begin{aligned}[b]
\frac{\partial f(V_n^m, \theta_n^p)}{\partial V_a}  \cong \frac{2(V_a -1)}{9} 
\end{aligned}
\end{equation}

\begin{equation}
\begin{aligned}[b]
\frac{\partial f(V_n^m, \theta_n^p)}{\partial \theta_a} \cong \frac{2\theta_a}{9} 
\end{aligned}
\end{equation}
Fig.~\ref{fig:Sensivity analysis illustration.} illustrates the sensitivity of VUF and MPVUR with respect to voltage magnitude and phase angle, showing that MPVUR exhibits bounded, near-constant magnitude sensitivities around nominal conditions, whereas VUF sensitivities vary with both magnitude and angle. Two key distinctions support the use of MPVUR as a surrogate penalty term in the IHL approach. First, at each node, MPVUR depends only on three voltage-magnitude variables, whereas VUF depends on six variables, including three phase angles embedded in trigonometric functions. Second, the derivatives of MPVUR are larger and admit nearly constant approximations across the three phases, which provides clearer descent directions and facilitates solver progress. This reduces the likelihood of degeneracy or near linear dependence among active constraints, thereby mitigating ill-conditioning in the KKT system and improving the reliability of Newton directions~\cite{NumOpt2006}. In contrast, VUF derivatives are smaller and more intricate, which can degrade numerical conditioning and hinder efficient optimization.

\section{Test cases}
\label{sec:test}
This section describes the test system, outlines the simulation setup, reports the results, and provides a comparative analysis with key observations.

All simulations are performed using the \texttt{PowerModelsDistribution.jl} package~\cite{FOBES2020106664}. The VU constraint and the penalty term are implemented externally in \texttt{Julia}~\cite{Bezanson2017} using \texttt{JuMP}~\cite{Dunning2017}, and integrated into the main OPF formulation. The solver employed in all cases is \texttt{Ipopt}~\cite{Ipopt2006},
and all simulations were performed on a laptop equipped with an Intel Core i7-1255U CPU @ 1.70 GHz, 16 GB of RAM. The developed code is publicly available on GitHub.\footnote{GitHub Repository: \url{https://github.com/alireza33zz/VU-ACOPF-v1.0.git}}

\begin{figure}[!t]
    \centering
    \includegraphics[width=2.8in]{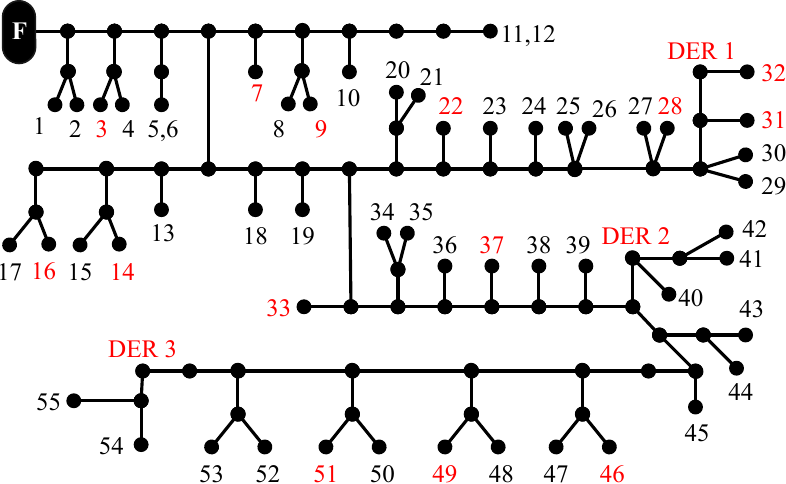}
    \caption{Modified European LV network schematic.}
    \label{fig:EU_LV_network_Schematics}
\end{figure}

The \textit{IEEE European low-voltage network} consists of 55 load buses and a substation feeder~\cite{ieeeEuLV}. A modified version of this network is used in the simulations as shown in Fig.~\ref{fig:EU_LV_network_Schematics}. Several rooftop PV panels, a three-phase solar power plant, and two inverter-based battery energy storage systems are incorporated into the base model, and an unbalanced load distribution is also applied. The details of the modified network are provided below.
The substation marginal energy cost is set to 1 €/kWh as a normalization reference for reporting DLMPs and CCoGs. 

Three DER units with reactive power support capabilities are integrated into the grid:
\begin{itemize}
    \item \textbf{DER2:} A solar power plant with zero marginal generation cost and a capacity of 54\,kVA, capable of providing up to 30\,kvar of reactive power support.
    \item \textbf{DER1 and DER3:} Inverter-based battery energy storage systems, each rated at 60\,kVA with a marginal dispatch cost of 1.1\,€/kWh (reflecting storage cycling costs~\cite{Battery}). Both units have a startup (activation) cost of 150\,€ and can provide up to 54\,kvar of reactive power support through their smart inverters~\cite{Battery2}.
\end{itemize}

Three-phase loads are connected at buses 13, 24, and 36, representing VU-sensitive equipment such as electric motors. Each load has a demand of 9\,kW with a lagging power factor of 0.93. The remaining load buses are equipped with single-phase loads ranging from 3.5 to 5.5\,kW, operating at an average power factor of 0.95.

Additionally, 14 buses host rooftop PV panel arrays (marked with red numbers), each rated at 7.5\,kVA and operating at unity power factor. The total renewable generation capacity is 159\,kW and 120\,kW of battery energy storage system.

The total active power demand in the network is 311.5\,kW. The phase-wise load distribution is as follows:
\begin{itemize}
    \item Phase $a$: 28.8\%
    \item Phase $b$: 33.2\%
    \item Phase $c$: 38.0\%
\end{itemize}

It should be noted that the official European LV benchmark comprises 907 nodes, which can lead to excessive runtimes and convergence issues with commercial solvers; therefore, the reduced-order model in~\cite{Ahmad} is used, preserving the electrical characteristics of the original network while reducing it to 117 nodes and substantially lowering computational complexity.

\subsection{Accuracy of MPVUR}
\label{Accuracy of MPVUR}
This subsection validates the magnitude-based approximation introduced in Section~\ref{MPVUR}. Since MPVUR is used as the penalty signal in IHL, its role is to provide a smooth and informative objective term that steers the optimizer toward low-unbalance operating points. In contrast, feasibility and grid-code compliance are still enforced through the hard VUF constraint. Therefore, the key requirement is not exact equality between MPVUR and VUF at every operating point, but rather that MPVUR reliably tracks VUF across buses and operating conditions and preserves the relative ranking of unbalance severity.

\begin{figure*}[!t]
\centering
    \includegraphics[width=0.80 \linewidth]{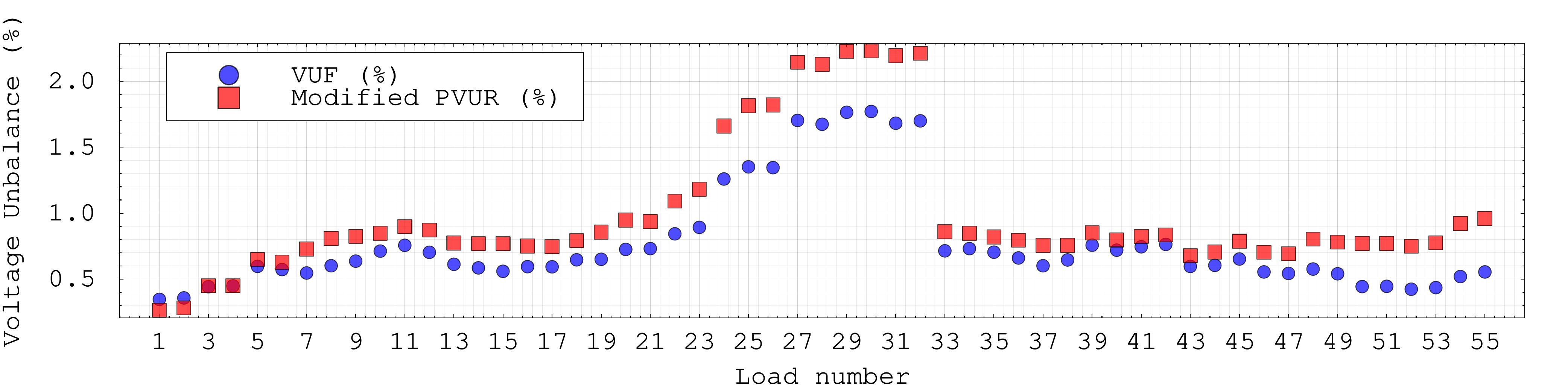}
\caption{Comparison of VUF and MPVUR in the IEEE European LV test network under the default OPF.}
\label{fig:PVUR accuracy1}
\end{figure*}

\begin{figure*}[!t]
\centering
    \includegraphics[width=0.80 \linewidth]{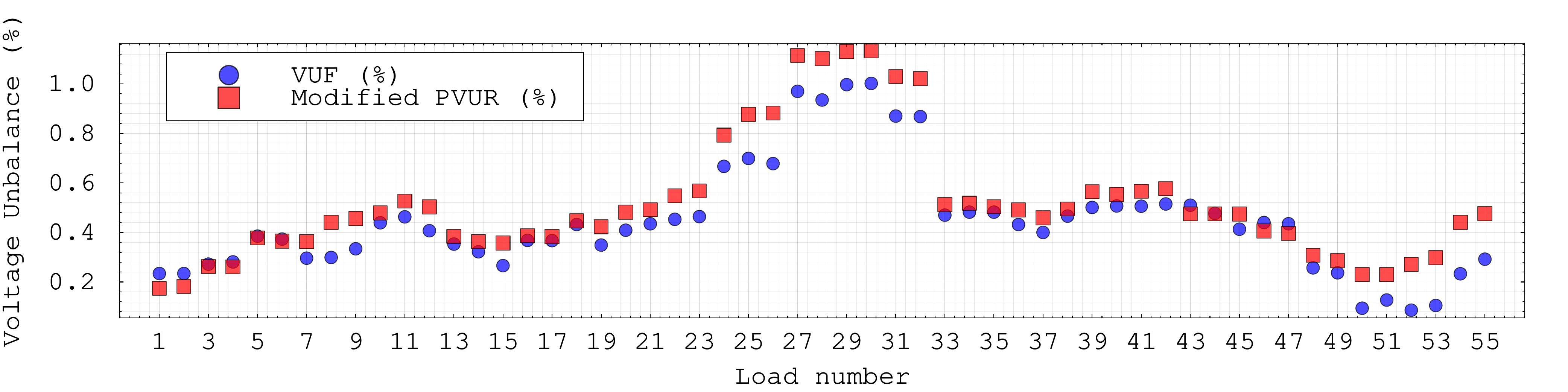}
\caption{Comparison of VUF and MPVUR in the IEEE European LV test network after VU mitigation via IHL.}
\label{fig:PVUR accuracy2}
\end{figure*}

Based on the described network, the VUF and MPVUR values obtained from the default OPF are shown in Fig.~\ref{fig:PVUR accuracy1}. The correlation between VUF and MPVUR is 0.979, with a mean error of 0.212\%. The corresponding results after VU mitigation using the IHL method are presented in Fig.~\ref{fig:PVUR accuracy2}, where the correlation between VUF and MPVUR is 0.962 and the mean error is 0.071\%. The reduction in mean error after mitigation indicates that the approximation becomes tighter in the operating region targeted by the optimization, i.e., when unbalance levels are suppressed toward the admissible range. The slight reduction in correlation is expected because the post-mitigation operating points occupy a narrower range of unbalance values, making the correlation metric more sensitive to small deviations.

Overall, both cases exhibit strong correlation and low mean error, confirming that MPVUR provides a suitable approximation of VUF for the intended purpose in IHL. In particular, MPVUR captures the spatial variation of unbalance across the feeder while avoiding the angle-dependent nonlinearities inherent to VUF, which supports its use as a computationally tractable penalty term without compromising feasibility.

\vspace{-15pt}

\subsection{Case studies}
In this section, five scenarios are considered:

\begin{itemize} 
\item \textbf{Default OPF:} VU is not considered in the optimization. 
\item \textbf{Hard limits:} VUF constraints (defined in~\eqref{eq:VUF2}) are imposed on all buses, i.e., VUF is used for $f$ in~\eqref{eq:V.U._constraint}. 
\item \textbf{Soft limits:} VUF is included as a penalty term in the objective function for all buses, i.e., VUF is used for $g$ in~\eqref{eq:social_welfare2}. 
\item \textbf{Hybrid limits:} VUF is included both as a penalty term in the objective function and as a constraint for all buses. 
\item \textbf{IHL:} Same as the hybrid limits case, but the penalty function is replaced with  MPVUR (defined in~\eqref{eq:MPVUR}).
\end{itemize}

To incorporate VU as a penalty term in the objective function, a weight \( \alpha \) must be selected. This coefficient represents the relative importance assigned to VU in the market-clearing objective. In the `soft limits' method, where VU is handled only through penalization, \( \alpha = 1 \) is selected to produce a visible suppression of VUF; for smaller values, the solution remains close to the default OPF. In the `hybrid limits' method and IHL, the VUF bound is enforced primarily by constraint~\eqref{eq:V.U._constraint}, and the penalty term is used mainly to provide a smooth guidance signal for the solver; therefore, a smaller value \( \alpha = 0.1 \) is adopted.

Tuning sensitivity differs across formulations. In the classic `hybrid limits' method, the same VU metric appears both as a hard constraint and as a penalty term, which can make convergence sensitive to the choice of \( \alpha \) and may lead to slow or failed solves for certain values. In the `soft limits' method, tuning is inherently a trade-off between suppressing VUF and avoiding unnecessary cost, and the required penalty level can be nontrivial and potentially location dependent. In contrast, IHL reduces this sensitivity by keeping feasibility enforcement in the explicit constraint and using a smoother proxy in the penalty term, which improves reliability while retaining the intended role of \( \alpha \) as a solver-guidance weight.

The results of these case studies are summarized in Table~\ref{tab:method_comparison}, reporting the convergence time, total generation cost, total losses, the highest VUF in the network, and whether feasibility with respect to the VUF bound is preserved. The hard limits case is omitted because enforcing VUF constraints at all buses yields an extremely tight feasible region and can introduce near linear dependence among active constraints, which leads to solver non-convergence; similar behavior has been reported in~\cite{churkin2024, MY-J1}. The key observations from Table~\ref{tab:method_comparison} are discussed next.

First, the `hybrid limits' method is markedly slower than IHL: `hybrid limits' requires 133.47~secs, whereas IHL converges in 17.50~secs (approximately an eight-fold speedup). Both methods enforce the same VUF bound, but IHL uses a smoother objective penalty, which yields faster and more reliable convergence.

Second, the cost of improving power quality and satisfying the VUF bound appears as an increase in generation cost from 461.52~€ (default OPF) to 477.62~€, i.e., +3.5\% over the reported interval. `Hybrid limits' and IHL produce the same generation cost (477.62~€) and nearly identical losses (5.54 vs. 5.56~kWh), indicating that replacing the VUF penalty with MPVUR does not materially change the operating point while improving computation.

Third, VU mitigation reduces total losses from 9.00~kWh (default OPF) to about 5.5--5.6~kWh (approximately 38\%), consistent with re-dispatch and reactive support that improve phase balance and reduce current-related losses.

Finally, the `soft limits' method converges quickly (11.17~s) and lowers the peak VUF (1.77\% to 1.17\%), but it does not preserve feasibility because no hard VUF bound is enforced. Achieving tighter reduction would require further penalty tuning, which can be nontrivial and bus dependent, and may move the solution away from the socioeconomic optimum. In contrast, hybrid limits and IHL both enforce the VUF bound (1.00\%), but only IHL combines feasibility with robust and efficient convergence in Table~\ref{tab:method_comparison}.

\begin{table*}[!t]
\renewcommand{\arraystretch}{1.3}
\centering
  \caption{Comparison of methods.}
\begin{tabular}{lcccccc}
\toprule
\textbf{Method} &
\makecell{\textbf{Convergence}\\\textbf{Time (sec)}} &
\makecell{\textbf{Generation}\\\textbf{Cost (€/h)}} &
\makecell{\textbf{Total}\\\textbf{losses (kWh)}} &
\makecell{\textbf{Highest}\\\textbf{VUF (\%)}} &
\makecell{\textbf{Respecting}\\\textbf{feasibility}} \\
\midrule
(a) Default OPF        & 13.66  & 461.52 & 9.00 & 1.77 & no \\
(b) Soft limits           & 11.17 & 472.66  & 5.62 & 1.17 & no  \\
(c) Hybrid limits          & 133.47 & 477.62 & 5.54 & 1.00 & yes \\
(d) Improved Hybrid limits & 17.50 & 477.62 & 5.56 & 1.00 & yes \\
\bottomrule
\end{tabular}
\label{tab:method_comparison}
\end{table*}

\subsection{Comparative assessment based on economic signals}
The comparison among VU-integration strategies is performed through the economic signals delivered by the market-clearing solution to consumers and generators. In particular, DLMP is used to characterize marginal incentives for consumption at each node and phase, while CCoG is used to quantify the economic impact of curtailment decisions on distributed generators. Together, these two quantities provide an interpretable link between power quality enforcement and market outcomes.

\begin{figure*}[!t]
    \centering
    \includegraphics[width=0.90 \linewidth]{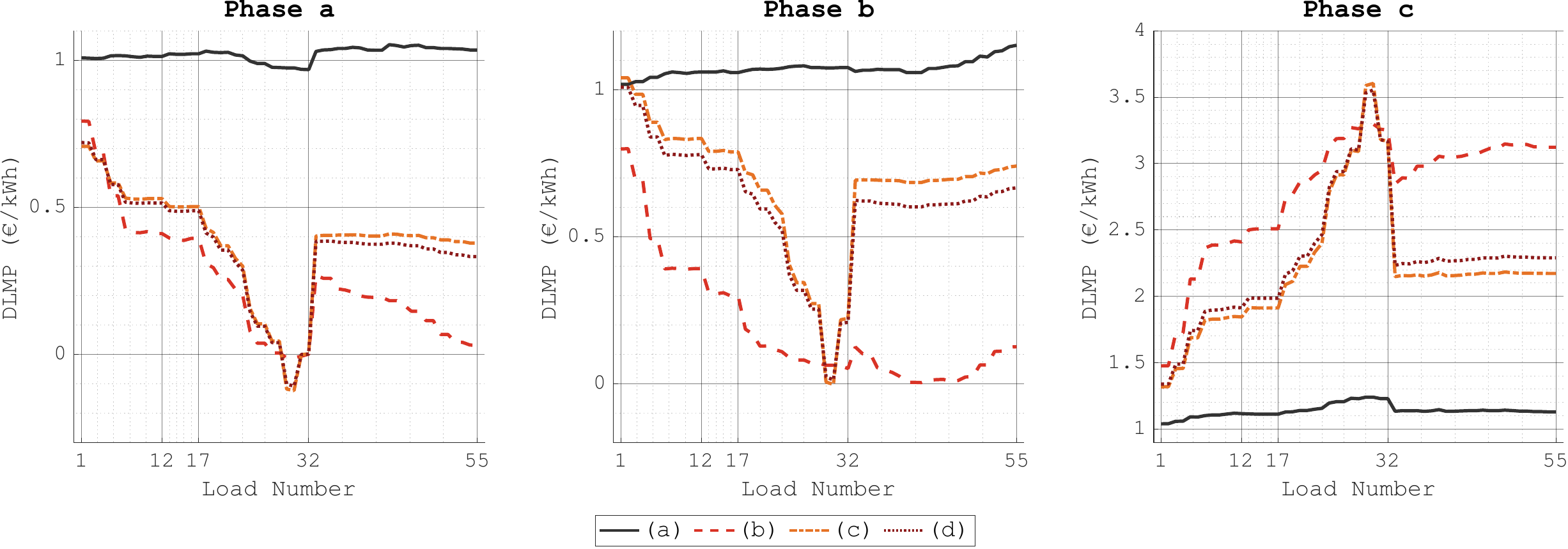}
    \caption{DLMP values of each load per phase for all case studies (Guide: (a):=Default OPF, (b):=Soft limits, (c):=Hybrid limits, and (d):=IHL).}
    \label{fig:combined_DLMP}
\end{figure*}

DLMP represents the marginal cost of serving an extra unit of demand at a given location and phase, and therefore captures how network constraints and power quality requirements propagate into nodal price signals. Fig.~\ref{fig:combined_DLMP} reports DLMP values for all loads across the three phases for the four case studies. The key observation is that the overall DLMP profiles remain consistent across the constraint-based methods, with the `hybrid limits' method and IHL exhibiting nearly indistinguishable curves on all phases. This indicates that replacing the VUF-based penalty by the MPVUR-based proxy in IHL does not distort the marginal incentives faced by consumers when the same VUF feasibility bound is enforced. By contrast, the soft limits case shows visible deviations in several locations, which is consistent with the absence of a hard feasibility requirement and the dependence of the solution on penalty tuning.

\begin{figure*}[!t]
    \centering
    \includegraphics[width=0.80 \linewidth]{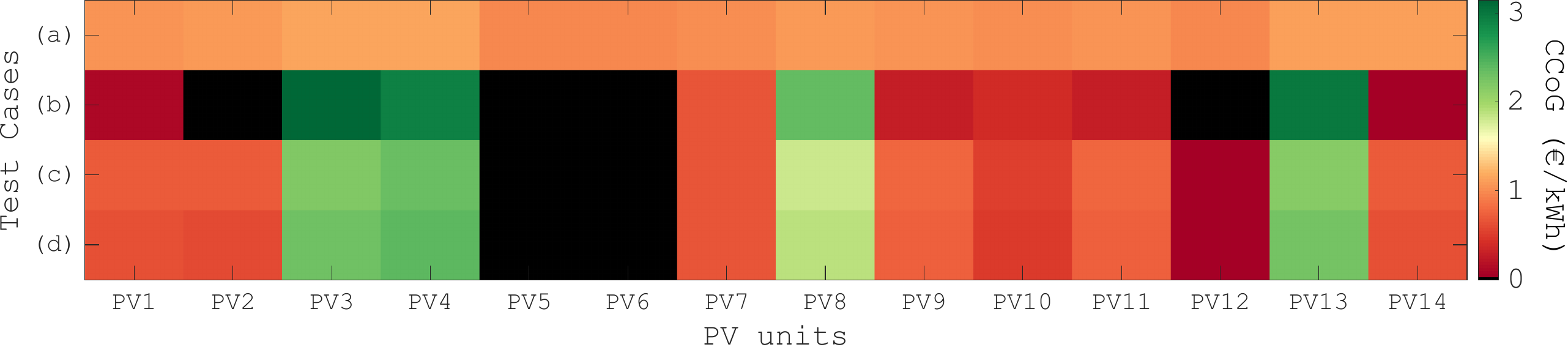}
    \caption{Curtailment cost of PV pannels (Guide: (a):=Default OPF, (b):=Soft limits, (c):=Hybrid limits, and (d):=IHL).}
    \label{fig:Dual_values_renewable_generators}
\end{figure*}

CCoG captures the opportunity cost associated with curtailing a generator, thereby indicating which units are economically most expensive to restrict and which curtailment actions are most consistent with network power quality objectives. Fig.~\ref{fig:Dual_values_renewable_generators} visualizes CCoG for PV units across the case studies. The `hybrid limits' method and IHL yield highly similar CCoG patterns, including the set of curtailed units (black cells) and the relative intensity of curtailment signals across PV locations. This agreement confirms that IHL preserves generator-side economic guidance while improving numerical behavior. The soft limits case can exhibit more pronounced variations, reflecting the fact that penalty-only formulations may trade feasibility against cost in a tuning-dependent manner. In the adopted color scheme, a shift toward green indicates that the unit contributes positively to VU mitigation, whereas a shift toward red indicates increased VU; black cells represent curtailment.

All convergent VU-aware formulations yield broadly consistent economic signals. Hybrid limits and IHL are nearly identical in power flow, DLMP, and CCoG, while outperforming soft limits by enforcing grid-code compliance with less tuning sensitivity. IHL further improves practicality by converging much faster and more reliably than classic hybrid limits, which remains tuning-sensitive and can be unstable.

\section{Conclusion}
\label{sec:conclusion}
The increasing penetration of single-phase PV and EV charging intensifies VU in distribution networks, making its explicit consideration essential for grid-code compliance and asset management. This paper investigated the integration of VU-aware three-phase AC-OPF into distribution-level market clearing and quantified the implications for feasibility, numerical performance, and economic signals.

A comparative assessment of hard, soft, hybrid, and IHL formulations on a European LV feeder showed that hard VUF enforcement across all buses can lead to non-convergence, while purely soft penalization does not guarantee feasibility and is highly sensitive to tuning. IHL addresses these limitations by combining hard VUF limits with a smooth MPVUR-based penalty, preserving compliance while improving numerical tractability. Results confirm that MPVUR closely approximates VUF, and that IHL produces DLMP and CCoG outcomes comparable to standard hybrid limits, with substantially faster and more reliable convergence.

Future work may address the tuning problem by defining an algorithm to consider monetization of VU or grid-code compliance, different surrogate functions with better numerical behavior, and linearization of the formulation to investigate possible solutions.


\section*{Acknowledgment}

This work was supported by MICIU/AEI/10.13039/501100011033 and ERDF/EU under grant PID2022-141609OB-I00, and by the Madrid Government (Comunidad de Madrid-Spain) under the Multiannual Agreement 2023-2026 with Universidad Politécnica de Madrid, `Line A - Emerging PIs' (grant 24-DWGG5L-33-SMHGZ1).
The work of Alireza Zabihi was supported by the 2023 FPI-UPM call for Predoctoral Contracts within the framework of the 2021-2023 State Plan for Scientific, Technical, and Innovative Research.

\ifCLASSOPTIONcaptionsoff
  \newpage
\fi

\bibliographystyle{IEEEtran}
\balance
\bibliography{Bibliography}
\end{document}